\documentclass[aps,pra,twocoloumn,reprint]{revtex4-2}
\usepackage[utf8]{inputenc}

\usepackage{graphicx,animate}
\usepackage{caption}
\usepackage{subcaption}
\usepackage{dcolumn}
\usepackage{bm}
\usepackage{physics}
\usepackage{ragged2e}
\usepackage{appendix}
\usepackage{amsmath}
\usepackage{amssymb}
\usepackage{listings}
\usepackage{comment}
\usepackage{xcolor}
\usepackage{tikz}
\usepackage{balance}
\allowdisplaybreaks

\def\x{3}
\def\y{1}

\begin{document}
	
	\title{Hysteresis and Self-Oscillations in an Artificial Memristive Quantum Neuron}
	\author{Finlay Potter, Alexandre Zagoskin, Sergey Savel'ev, and Alexander G Balanov}
	\affiliation{Department of Physics, Loughborough University, Loughborough, LE11 3TU, United Kingdom}
	

	\begin{abstract}
		We theoretically study an artificial neuron circuit containing a quantum memristor in the presence of relaxation and dephasing. The charge transport in the quantum element is realized via  tunneling of a charge through a quantum particle which shuttles between two terminals --  a functionality reminiscent of classical diffusive memristors.  We demonstrate that this physical principle enables hysteretic  behavior of the current-voltage characteristics of the quantum device. In addition, being used in artificial neural circuit, the quantum switcher  is able  to generate self-sustained current oscillations. Our analysis reveals that these self-oscillations are triggered only in quantum regime with a moderate rate of relaxation, and cannot exist either in a purely coherent regime or at a very high decoherence. We investigate the hysteresis and instability leading to the onset of current self-oscillations and analyze their properties depending on the circuit parameters. Our results provide a generic approach to the use of quantum regimes for controlling  hysteresis and generating self-oscillations.
	\end{abstract}
	
	\maketitle
	\section{Introduction.}
	
	Memristive technologies have attracted increasing research attention due to their potential in computing, memory storage, artificial intelligence hardware, and communication applications \cite{Lanza2022}. These technologies rely on a specific class of electronic devices conventionally called ``memristors", which are essentially switchers with memory \cite{Kumar2022, Xiao2023aa}. While the concept of memristors dates back to 1971 \cite{Chua1976, Chua2012}, their experimental realization only occurred in 2007, requiring the adoption of novel physical principles and materials \cite{Strukov2008}.
	
	A defining characteristic of memristors is their ability to ``remember" the history of current through them, altering their conductance based on this data. Such a behavior is reflected in a distinctive pinched hysteretic loop in their input-output dependence. This unique property allows memristors to perform both memory and processing functions within a single device. Furthermore, features such as controllable volatility and stochasticity, coupled with high integration density, position memristive devices as highly promising for the future of electronics and as contributors to extending the life of Moore's Law \cite{DMarkovi}. Remarkably, the generic paradigms of memristive devices  \cite{Chua1976} extends beyond the realm of electronics. Its implementation can be explored in diverse fields, including spintronics or photonics \cite{Chanthbouala2012, Spagnolo2022, Qin2023}.
	
	The current surge in memristor research is largely driven by their potential applications in neuromorphic computing \cite{Christensen:2022aa}. This innovative approach offers an alternative to the traditional von Neumann computing architecture, aiming to overcome its primary weakness, the "von Neumann bottleneck": the slowdown caused by the exchange of data between separate storage and processing units \cite{Beyond2020, Sebastian2017,Horowitz2014,Liu2020}.
	In contrast, neuromorphic computing simulates the brain's structure by integrating data storage and processing within interconnected functional elements called artificial neurons and artificial synapses. These neurons, functioning as circuits, generate nonlinear outputs in response to integrated inputs. Memristor devices, with the capability to both memorize and process signals within the same unit, offer an opportune realization of efficient components for neuromorphic systems \cite{AAkther,WOJTUSIAK2021,ZWang, WXu, Park2022, Rao:2023aa}. 
	
	Recently, the notion of quantum memristors \cite{Pfeiffer2016,Shevchenko:2016aa,Salmilehto2017,Sanz2018} has appeared, presenting a complement to quantum counterparts of other classical circuit elements  \cite{Zagoskin2011}. This conceptual framework aims to harness quantum mechanical principles to not only enhance the performance and capabilities of memristive devices but also to open new pathways for information processing  \cite{Markovic:2020}. Moreover, it offers valuable insights into the role of quantum phenomena in highly dense memristive circuits, where the impact of quantum effects becomes pronounced due to the small size of individual memristors. Despite some recent development \cite{Norambuena:2022aa, Huggins2023, Tang:2023aa, Stremoukhov:2023aa}  and  preliminary experimental results \cite{Spagnolo2022}, this area remains largely unexplored. Among  key remaining questions are the identification of physical principles for practical implementation, an assessment of quantum memristor performance in both classical and quantum circuits, and the development of effective approaches to control their characteristics. 
	
	In this work, we explore a possibility to realize the transport principle which is inherent in a diffusive memristor \cite{Jiang:2017aa,AAkther}, wherein large particles provide a preferable path for electrons when tunnelling across spatial regions. Our proposal introduces a simplified quantum coherent system,  employing  a charged quantum particle acting as moving island, through which electrons tunnel between two terminals.  Within our study, this quantum device assumes the role of an active component within a classical circuit, typically  serving as the memristive artificial neuron in neuromorphic systems (see Fig. \ref{DeviceAndCircuit}(a)) \cite{Pickett:2013aa,ZWang}.
	
	We demonstrate that an artificial neuron incorporating the proposed quantum device exhibits hysteresis in the output voltage under changes in the voltage applied to the circuit. Furthermore, this hysteresis persists both in the coherent states of the quantum device and when the quantum state is allowed to fully thermalize. Additionally, the artificial neuron is capable of showing self-sustained oscillations under moderate levels of environmental dissipation by the quantum system. Analysis reveals the robustness of both phenomena, which exist across a range of system parameter values.

	\section{Model for Artificial Neuron with a Memristive Quantum Element.}
	\label{sec:model}
	\begin{figure}
		\centering
		\begin{subfigure}[b]{0.46\textwidth}
			\centering
			\includegraphics[width=\linewidth]{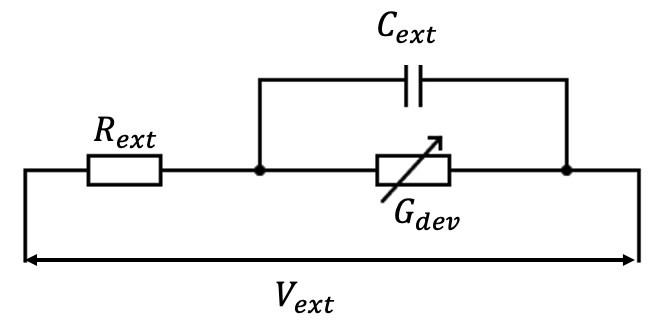}
			\caption{}
			\label{Circuit}
		\end{subfigure}
		\begin{subfigure}[b]{0.49\textwidth}
			\centering
			\begin{tikzpicture}[scale=0.85, every node/.style={scale=1}]
				\fill[blue!60!white] (0,8-\y) rectangle (1,11-\y);
				\fill[red!60!white] (8,8-\y) rectangle (9,11-\y);
				
				\node[] at (4+\x*0.5,10.5-\y) {$e^-$};
				\draw (4+\x*0.5,10.5-\y) circle (0.3cm);
				\draw[thick, ->] (\x,8.5-\y) -- (\x+1.2,8.5-\y);
				\draw[thick, ->] (\x,8.5-\y) -- (\x-1.2,8.5-\y);
				\shade[inner color=black,outer color=white] (\x,2.9+6.5-\y) circle (0.7cm);
				
				\draw[thick] (0.5+0.5*\x,4.2+6.7-\y) parabola (1,3.33666002653 +6.7-\y);
				\draw[thick,->] (0.5+0.5*\x,4.2+6.7-\y) parabola (\x,3.33666002653 +6.7-\y);
				\draw[thick] (4+\x*0.5,4.2+6.7-\y) parabola (\x,3.33666002653+6.7-\y);
				\draw[thick,->] (4+\x*0.5,4.2+6.7-\y) parabola (8,3.33666002653+6.7-\y);
				
				\fill[blue!60!white] (0,0) rectangle (1,6);
				\fill[red!60!white] (8,0) rectangle (9,6);
				\draw[thick, ->] (1,-0.1) -- (8.,-0.1);
				\draw (\x,0) parabola (\x+1.7,6);
				\draw (\x,0) parabola (\x-1.7,6);

				\draw (\x-1.2,1.05)..controls (\x-0.5,0.95) and (\x-0.2,2.05) .. (\x,2);
				\draw (\x,2) .. controls (\x+0.2,2.05) and (\x+0.5,0.95) .. (\x+1.2,1.05);
				\draw (\x-1.2,1.04) -- (\x+1.2,1.04);
				
				\draw (\x -1.5,3.2)..controls (\x-1,3.3) and (\x-0.7,2).. (\x,3.2);
				\draw (\x,3.2).. controls (\x+0.7,4.4) and (\x+1,3.2) .. (\x+1.5,3.3);
				\draw (\x -1.5,3.25) -- (\x+1.5,3.25);
				
				\draw[thick,->] (4.5,-0.1) -- (4.5,6);

				\node[] at (\x,-0.3) {$x_0$};
				\node[] at (4.5,-0.3) {$0$};
				\node[] at (8.35,-0.15) {$x$};
				\node[] at (1,-0.4) {$-L$};
				\node[] at (8,-0.4) {$L$};
				\node[] at (4.5,6.3) {$E$};
				\node[] at (4.9,1.05) {$\frac{\hbar\Omega_p}{2}$};
				\node[] at (4.9,3.2) {$\frac{3\hbar\Omega_p}{2}$};
				\node[] at (-0.3,3) {$-$};
				\node[] at (9.3,3) {$+$};
				\node[] at (\x - 1.3,1.3) {$\psi_0$};
				\node[] at (\x - 1.57,3.45) {$\psi_1$};
			\end{tikzpicture}
			\caption{}
			\label{Device}
		\end{subfigure}
		
		\caption{(a) Artificial neuron circuit with $G_{dev}$ representing the conductance of the device in (b). (b) Schematic of the quantum memristive device: first two energy levels and the related wave functions of a quantum particle in a harmonic potential between two contacts denoted as ``-" and ``+".}
		\label{DeviceAndCircuit}
	\end{figure}
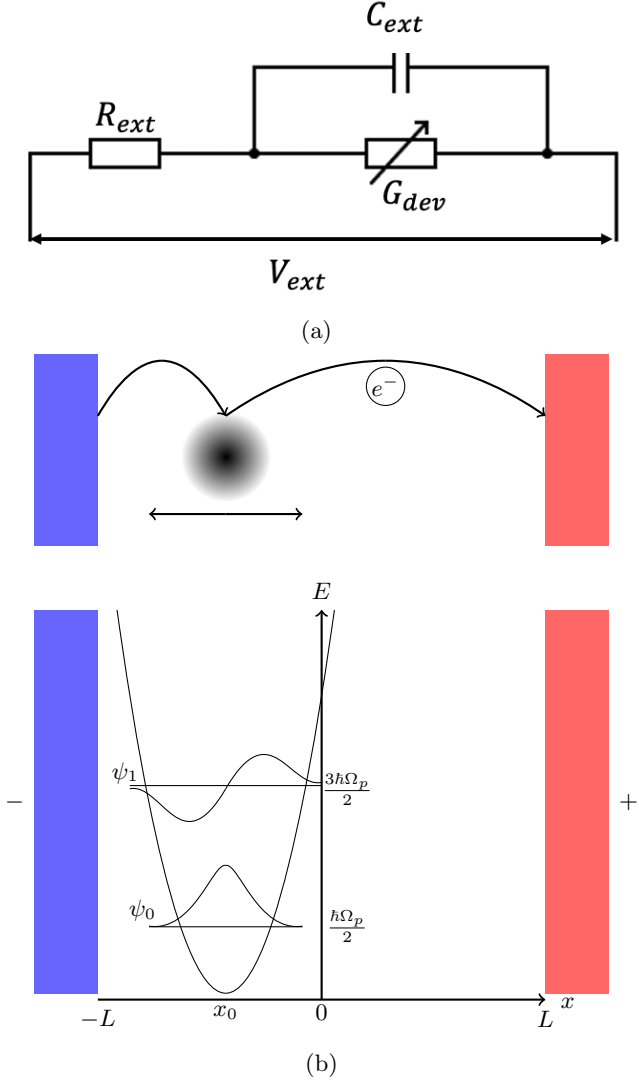
	The system under study is schematically presented  in Fig. \ref{DeviceAndCircuit}. The  charge transport in the quantum device is realized  by sequential tunneling of electrons from one of the contacts to a charged quantum particle moving under a potential, which can be locally approximated as a simple harmonic potential, and then to another contact, see Fig. \ref{DeviceAndCircuit}(b). 
	Both the rate of tunneling and the particle state are affected by the electric field generated by the applied voltage. Furthermore, the  charge tunneling rate then depends on the particle state  (i.e. the conductance of the device). Consequetially, the particle state therefore depends on the history of tunneling, thus providing a mechanism for memristive properties.
	
	Practical realization of such a regime  still requires an experimental  implementation. This may be possible by utilizing approaches similar to ones proposed in  Refs \cite{JKettler2021, kunne2023spinbus, Yi2016}.
	
	A circuit diagram of the artificial neuron, which we analyze in this paper  is presented in Fig. \ref{DeviceAndCircuit}(a). Here $G_{dev}$ represents the conductance of the device in \ref{DeviceAndCircuit}(b), $C_{ext}$ and $R_{ext}$ denote the capacitance  and the resistance of the external circuit, and $V_{ext}$ is an applied voltage bias. In this configuration, $V_{ext}$  serves as an integral neural input, the voltage drop across the quantum element  $V_m$ provides  the neuron output, and $\tau_c=C_{ext}R_{ext}$ defines the characteristic timescale of the circuit.
	
	To study the conducting  properties of the quantum device shown in Fig. \ref{DeviceAndCircuit}(b) we consider its Hamiltonian as
	
	\begin{equation}
		\label{eq:H}
		\mathcal{H} = \frac{-\hbar^2}{2m}\frac{\partial^2}{\partial x^2} + m\Omega_p^2(x-x_0)^2 + \frac{qV_m}{2L}x.
	\end{equation}
	Here, $m$ and $q$ are the quantum particle's mass and charge, respectively, $x$ is its position, $x_0$ and $\Omega_p$ is the position of the minimum and the angular frequency of the simple harmonic potential, and $2L$ is the distance between the terminals. The final term in this Hamiltonian represents the action of the electric field between the two terminals, where $V_m$ is the voltage dropped across the memristor. This action is assumed to be homogeneous on the charged particle.
	
	By making the substitution $\tilde{x} = x-x_0 +qV_m/(2m\Omega_p^2 L)$  the Hamiltonian becomes
	
	\begin{equation}
		\mathcal{H} = -\frac{\hbar^2}{2m}\frac{\partial^2}{\partial\Tilde{x}^2} + \frac{1}{2}m\Omega_p^2\Tilde{x}^2 - i\hbar\frac{q}{2m\Omega_p^2L}\frac{dV_m}{dt}\frac{\partial}{\partial\Tilde{x}}, \label{Hamiltonian}
	\end{equation}
	see Appendix \ref{appendix:Hamiltonian} for details. The Hamiltonian (\ref{Hamiltonian}) is characterized by the set of harmonic oscillator eigenfunctions
	
	\begin{equation}
		\psi_n(\tilde{x}) = \frac{1}{\sqrt{2^nn!}}\left(\frac{1}{\pi l^2}\right)^{\frac{1}{4}}e^{-\frac{\Tilde{x}^2}{2l^2}}H_n\left(\frac{\Tilde{x}}{l}\right),
	\end{equation}
	where  $n$ is a quantum number,  $l = \sqrt{\hbar/(m\Omega_p)}$  is the characteristic length of the quantum harmonic oscillator and $H_n(u)$ is the $n^{th}$ Hermite polynomial for the argument $u$. Assuming the complete potential is sufficently anharmonic that the second excited state differs from one of a harmonic potential  by $\hbar\Delta\Omega$, and that $\Omega_P+\Delta \Omega\gg\Omega_p$, we limit our consideration to the first two energy levels.
	
	Within this approximation, the Hamiltonian matrix has the form
	\begin{equation}
		\mathcal{H} =\hbar
		\begin{pmatrix}
			\frac{1}{2}\Omega_p&-\frac{1}{2\sqrt{2}}\frac{q}{m\Omega_p^2 l L}\frac{dV_m}{dt}i\\
			\frac{1}{2\sqrt{2}}\frac{q}{m\Omega_p^2 l L}\frac{dV_m}{dt}i & \frac{3}{2}\Omega_p 
		\end{pmatrix}.
		\label{HamiltonianN}
	\end{equation}

	To describe the quantum state of the device in Fig. \ref{DeviceAndCircuit}(b)  we introduce the density matrix $\rho$:
	\begin{equation}
		\rho = \frac{1}{2}\begin{pmatrix}1+Z & X-iY\\X+iY&1-Z\end{pmatrix}, 
	\end{equation}
	where $X$, $Y$, $Z$  are the components of the Bloch vector describing this state.
	
	Substituting $\rho$ into the Liouville-von Neumann equation ${\displaystyle i\hbar \dot{\rho}=[\mathcal{H},\rho ]}$,
	and parameterizing a generic dissipation by relaxation and pure dephasing rates $\gamma_T$ and $\gamma_\varphi$ respectively \cite{SSaveliev, Zagoskin2011}, yields the dynamics equations for $X$, $Y$, $Z$:
	
	\begin{eqnarray}
		\label{eq:xyz}
		\frac{dX}{dt} &=& \Omega_p Y + \frac{1}{\sqrt{2}}\frac{q}{m\Omega_p^2 l L}\frac{dV_m}{dt}Z - \gamma_{\varphi}X, \nonumber \\
		\frac{dY}{dt} &=& -\Omega_p X - \gamma_{\varphi}Y,\\
		\frac{dZ}{dt}&=& -\frac{1}{\sqrt{2}}\frac{q}{m\Omega_p^2 l L}\frac{dV_m}{dt}X - \gamma_T(Z-Z_T).\nonumber
	\end{eqnarray}
	Here, $Z_T=\tanh[\hbar\Omega_p/(2k_B T)]$ is the thermal equilibrium value of $Z$  at temperature $T$ (see Appendix \ref{appendix:Therm})\cite{Zagoskin2011}. Note that dynamics within the Bloch sphere requires $\gamma_T<2\gamma_\varphi$ \cite{Bellac2006, Zagoskin2011} (see Appendix \ref{appendix:Dephasing}), which is a natural constraint for the relaxation parameters. Hereafter we set $\gamma_\varphi=\gamma$ and $\gamma_T=\alpha\gamma$, where $\alpha\in [0,2]$.
	
	We should note that the two-level approximation is only suitable if $\Delta\Omega\gg\Delta\epsilon$, where $\Delta\epsilon=\gamma+\alpha\gamma$, as this prevents the dissipation from producing resonance with higher energy levels.
	
	Previously, it has been estimated \cite{Saveliev2013} that tunneling of the electrons between the terminals through a classical shuttling particle produces conductance $G=\sech{(x/\lambda)}/R_0$, where $x$ is  the position of  shuttling particle,  $\lambda$ characterizes the tunneling length for the electrons,  and $R_0$ is the minimal resistance of the device. Therefore,  under assumption of  negligible  higher-order cumulants, the conductance in the case of a quantum shuttling particles  is
	
	\begin{equation}
		\begin{gathered}
			G_{dev}(V_m)=\langle G\rangle = \frac{1}{R_0}\Tr{\rho\left[\sech{\left(\frac{x}{\lambda}\right)}\right]}= \\\frac{1}{R_0}\Tr{\rho\left[\sech{\left(\frac{\Tilde{x} + x_0 -\frac{q}{2m\Omega_p^2L}V_m}{\lambda}\right)}\right]}.
		\end{gathered}
	\end{equation}
	Applying the Kirchoff's circuit law for currents to the circuit in Fig. \ref{DeviceAndCircuit}(a) gives the equation describing the artificial neuron output $V_m$
	\begin{equation}
		\label{eq:V}
		\tau_c \frac{dV_m}{dt}= V_{ext} - (1 + R_{ext}G_{dev}(V_m))V_m.
	\end{equation}
	Equations (\ref{eq:xyz})-(\ref{eq:V}) constitute the mathematical model for the artificial neuron under investigation.
	
	\begin{table}[h]
		\renewcommand{\arraystretch}{1.5}
		\begin{tabular}{c|c}
			Dimensionless Quantity & Definition                               \\ \hline
			$\tilde{t}$            & $\frac{t}{\tau_c}$                       \\
			$\Omega$               & $\Omega_p \tau_c$                        \\
			$\Gamma$               & $\gamma\tau_c$                           \\
			$V$                    & $\frac{q}{2\sqrt{2}m\Omega_p^2Ll}V_m$     \\
			
			$V_n$                  & $\frac{q}{2\sqrt{2}m\Omega_p^2Ll}V_{ext}$ \\
			$R_n$                  & $\frac{R_{ext}}{R_0}$                   
		\end{tabular}
		\caption{Dimensionless quantities and their definitions in scaled parameters of the system presented in Fig.\ref{DeviceAndCircuit}.}
		\label{Table}
	\end{table}
	
	For convenience, we will we make a transformation to dimensionless quantities given in TABLE \ref{Table}. The resulting model after these transformations reads:
	
	\begin{eqnarray}
		\label{eq:nmod0}
		\dot{Z} &=& -2\dot{V}X-\alpha\Gamma(Z-Z_T), \nonumber \\
		\dot{X} &=& \Omega Y +2\dot{V}Z - \Gamma X,  \\
		\dot{Y} &=& -\Omega X -\Gamma Y, \nonumber \\
		\dot{V} &=& V_n - \left(1+R_n\Tr{\rho\left[\sech{\left(\frac{\tilde{x}+x_0-l\sqrt{2}V}{\lambda}\right)}\right]}\right)V. \nonumber
	\end{eqnarray}
	Here, overdot means a derivative with respect to $\Tilde{t}$. Introducing $x_V = x_0-l\sqrt{2}V$ and, noticing that the conductance matrix elements with indices $i_,j$ are
	$[\sech((\tilde{x}+x_V)/ \lambda)]_{i,j}= \bra{\psi_i}\sech{(\tilde{x}+x_V)/{ \lambda})\ket{\psi_j}}$, we expand
	
	\begin{equation}
		\begin{gathered}    
			\Tr{\rho\left[\sech{\left(\frac{\tilde{x}+x_V}{\lambda}\right)}\right]} = \\\frac{1}{2}(1+Z)F_{0,0}(x_V) + XF_{0,1}(x_V)+\frac{1}{2}(1-Z)F_{1,1}(x_V) \nonumber
		\end{gathered}
	\end{equation}
	with
	\begin{equation}
		\begin{gathered}
			F_{i,j}(x_V) = \\\frac{1}{\sqrt{2^{i+j}i!j!}}\frac{1}{l\sqrt{\pi}}\int_{-\infty}^\infty e^{-\frac{\Tilde{x}^2}{l^2}}H_i\left(\frac{\tilde{x}}{l}\right)H_j\left(\frac{\Tilde{x}}{l}\right)\sech{\left(\frac{\tilde{x}+x_V}{\lambda}\right)}d\Tilde{x}. \nonumber
		\end{gathered}
	\end{equation}
	
	Example plots of these functions are presented in Appendix \ref{Appendix:F}.
	
	We can now define a dimensionless conductance $G(V)$ as being $\Tr{\rho[\sech{((\tilde{x}+x_V)/\lambda)}]}$ and a corresponding dimensionless current as $I(V) = G(V)V$.
	
	The system (\ref{eq:nmod0}) can then be written explicitly
	\begin{equation}
		\begin{aligned}
			\label{eq:nmod}
			\dot{Z} =& -2\dot{V}X-\alpha\Gamma(Z-Z_T), \\
			\dot{X} =& \Omega Y +2\dot{V}Z - \Gamma X,  \\
			\dot{Y} =& -\Omega X -\Gamma Y,\\
			\dot{V} =& V_n - \Bigl\{1+\frac{R_n}{2}[(1+Z)F_{0,0}(x_V)+\\& 2XF_{0,1}(x_V)+(1-Z)F_{1,1}(x_V)]\Bigr\}V.
		\end{aligned}
	\end{equation} Unless it is specified differently, we set the parameter values to $l/L= 0.5$, $x_0/L = 0.8$, $\lambda/L = 0.13$, $\alpha=1$, and $\Omega=7$.
	
	\section{Static Input-Output Characteristics}
	\begin{figure}
		\centering
		\begin{subfigure}[b]{0.45\textwidth}
			\centering
			\includegraphics[width=\linewidth]{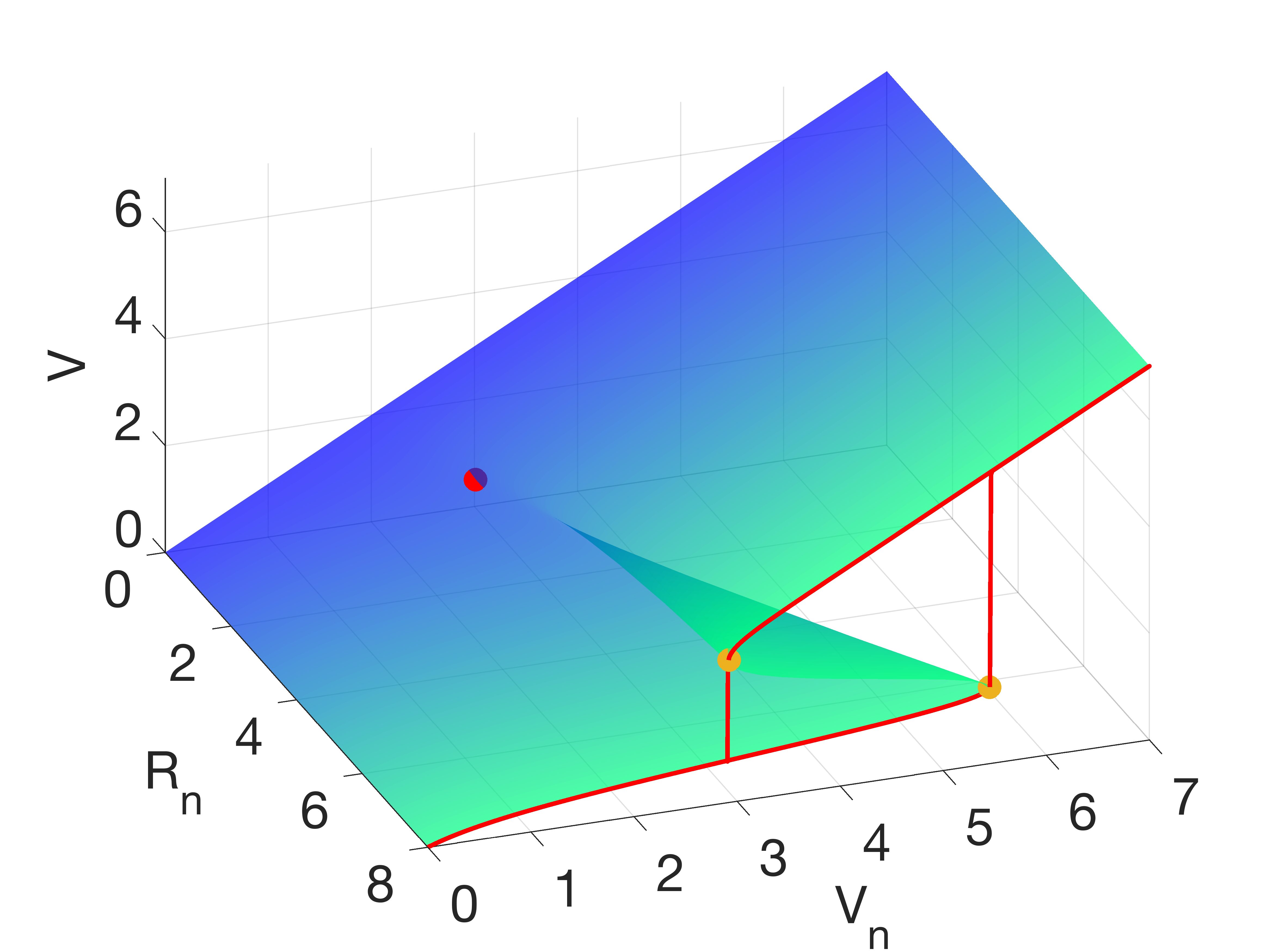}
			\caption{}
			\label{2LEqPoints,A}
		\end{subfigure}
		\begin{subfigure}[b]{0.45\textwidth}
			\centering
			\includegraphics[width=\linewidth]{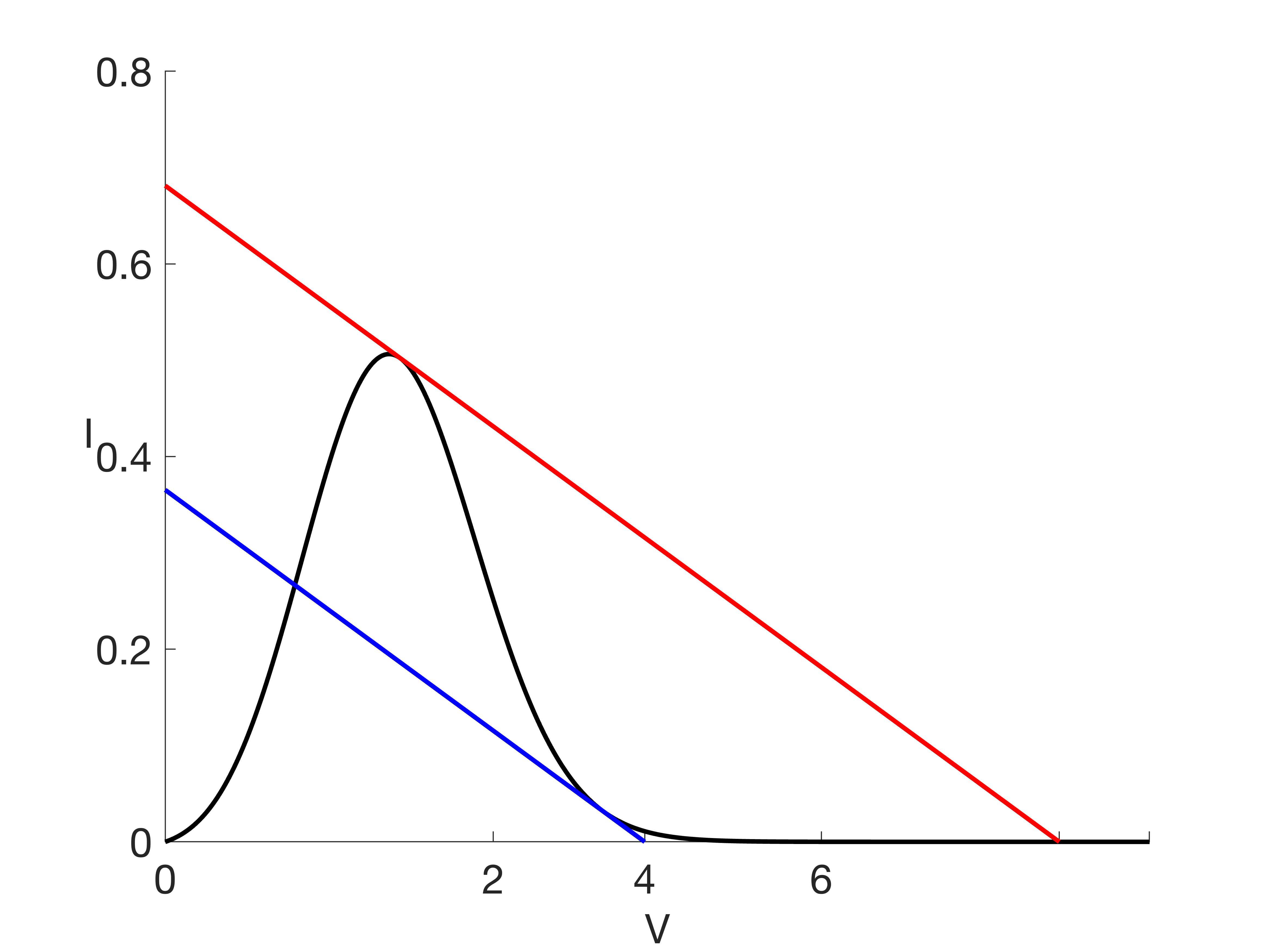}
			\caption{}
			\label{2LEqPoints,LL}
		\end{subfigure}
		\begin{subfigure}[b]{0.45\textwidth}
			\centering
			\includegraphics[width=\linewidth]{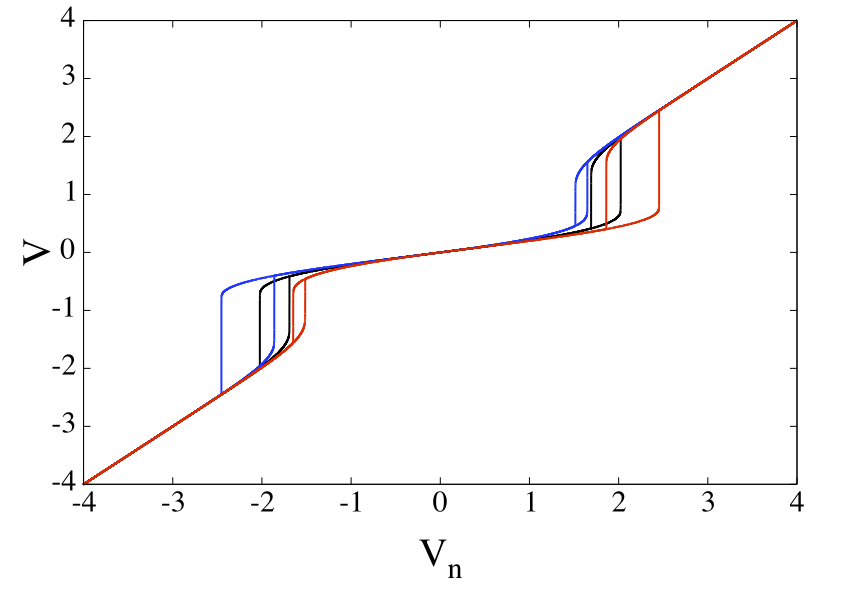}
			\caption{}
			\label{2LEqPoints,B}
		\end{subfigure}
		\caption{Hysteresis demonstrated in the artificial neuron circuit with $Z_T = 1$ (zero temperature limit). a)  Surface of equilibrium points describing the dependency $V^*=V^*(R_n,V_n)$ with a red curve showing the $V_n$ ramp calculated for $R_n=8$ with $\Gamma = 1$ and $\alpha=1$. b) (black) Current, I, as a function of voltage calculated as $I=G_{dev}(V)V$ with load lines for $R_n=8$ with $V_n = V_n^{(1)}=5.4501$ (red) and $V_n = V_n^{(1)}=2.9232$ (blue). c) Two regions of hysteresis occurring for $R_n = 10$ with $x_0=0$ (black), $x_0=0.1$ (red), and $x_0=-0.1$ (blue).}
		\label{2LEqPoints}
	\end{figure}
	
	First, we study the constant (DC) response $V$  of the artificial neuron with a constant input  $V_n$. For this aim we find the equilibria of (\ref{eq:nmod}) for a range of $V_n$ and $R_n$.  For non-zero $\alpha$ and $\Gamma$ the system (\ref{eq:nmod}) is in equilibrium if 
	\begin{equation}
		\begin{aligned}
			\label{eq:eq}
			X=& 0, \\
			Y=& 0,  \\
			Z=& Z_T, \\
			V=& V^*,
		\end{aligned}
	\end{equation}
	provided that $V^*$ satisfies
	\begin{equation}
		f(V^*) = 0,
	\end{equation}
	where $f(V)$ is defined as
	\begin{equation}
		\begin{gathered}
			\label{eq:f(V)}
			f(V) = V_n - \left\{1+\frac{R_n}{2}\left[(1+Z_T)F_{0,0}(x_0 - l\sqrt{2}V)\right.\right.\\\left.\left.+(1-Z_T)F_{1,1}(x_0 - l\sqrt{2}V)\right]\right\}V.
		\end{gathered}
	\end{equation}
	Hence, the DC-component of the neuron response $V$ depends on the circuit parameters $V_n$ and $R_n$,  electron tunneling parameters $x_0$, $l$  and $\lambda$ and the thermal state $Z_T$. The surface of equilibrium states $V^*=V^*(R_n, V_n)$ with $Z_T = 1$ and the tunneling parameters given in Section \ref{sec:model} is presented in Fig. \ref{2LEqPoints}(a). The equilibria depend only on the diagonal elements of the matrix $\left[\sech((\tilde{x}+x_V)/\lambda)\right]$, and therefore do not involve any quantum coherent processes.
	
	For small $R_n<1.77$ the equilibrium voltage $V^*$ almost linearly depends on the input $V_n$. However for $R_n>1.77$, the dependence $V^*=V^*(R_n, V_n)$  starts to demonstrate a fold, thus implying coexistence of three equilibrium values of $V$ for a single value of $V_n$, two stable and one unstable for a fixed $X$, $Y$, and $Z$. These coexisting equilibria produce a hysteretic response in the voltage across the memristor, $V$, which for $R_n=8$ is shown by a solid red line in Fig.\ref{2LEqPoints}(a). With increase of $V_n$ the response $V$ gradually increases, but when $V_n$ reaches the value $\approx 5.3$,  $V$ demonstrates a jump. Further increase of $V_n$ leads again to smooth and monotonic increase of $V$. If now $V_n$ decreases we observe that $V$ decreases by a jump this time at $V_n\approx 3$ followed by the further monotonic decrease. These hysteretic jumps occurring at different $V_n$ are associated with the onset of saddle-node (fold) bifurcations \cite{SStrogatz} indicated in  Fig. \ref{2LEqPoints}(a) by orange circles. This kind of bifurcation is associated with an instability, in the result of which an unstable (saddle) equilibrium in the phase space collides with a stable one.  The collision leads to disappearance of the involved equilibria leaving only a sole stable equilibrium. 
	
	The physical mechanism of the observed hysteresis  is similar to one discussed in Ref.\cite{Saveliev2013}. The conductivity between the terminals, realized by the electron tunneling to and from the shuttling particle, depends on the bias voltage non-monotonically, as seen in the current voltage characteristic $I(V) = G(V)V$ (the black line in Fig.\ref{2LEqPoints}(b)). For a large enough $R_n$, there will be a range of $V_n$ where the load line, joining points $R_n V_n$ and $V_n$, intersects the $I(V)$ curve in three places, indicating hysteresis.
	
	Therefore, the same conductance, $G$, could be realized for  different values of voltage, $V$.  If such a device is included in the circuit shown in Fig. \ref{DeviceAndCircuit}(a)  then in the DC-current regime ($\dot{V}=0$)  for large enough resistances $R_n$, the external input voltage $V_n$ could lead to several different values of $V$. The coexistence of three different options for $V$ to compensate the same value of $V_n$  constitutes a mechanism for the hysteresis.
	
	This explanation for the hysteresis is valid for both positive and negative $V_n$ when calculating the dependency $V^*(V_n)$ for a sufficiently large value of $R_n$, and is confirmed by numerical analysis of  Eq. (\ref{eq:nmod}); symmetric  hysteresis loops in the dependence $V$ as $V_n$ is varied are obtained with $x_0=0$ shown in  Fig. \ref{2LEqPoints}(c) in black.  However, when the position of the harmonic potential minimum changes, i.e.  $x_0\ne 0$, the loops for positive and negative $V_n$ become asymmetric; one of them shrinks while another widens (see orange and blue curves  in \ref{2LEqPoints}(c)), since in this case, for the same voltage magnitude the tunneling rate in one of the directions become greater than in another.
	
	Remarkably, the hysteretic behavior exists for any $Z_T$ because both $F_{0,0}(x_0-l\sqrt{2}V)$ and $F_{1,1}(x_0-l\sqrt{2}V)$ have maxima in dependence on $V$, see Fig. \ref{FijPlots} in the Appendix \ref{Appendix:F}.  Therefore, the existence of memristive properties of the device does not depend on temperature and is common both for classical and quantum shuttling particle.
	
	By analyzing the function $f(V)$ given as our condition for equilibrium in Eq.(\ref{eq:f(V)}), we can determine the critical values of $R_n$, $V_n$, and $V^*$ where $V^*(R_n,V_n)$ becomes multi-valued. This is the point at which the cusp bifurcation occurs, beyond which we expect to be able to observe hysteresis in the DC voltage characteristics. For the parameters used in Fig.\ref{2LEqPoints}(a) the coordinates of this are found to be $V^{(cusp)}\approx 1.8866$, $R_n^{(cusp)} \approx 1.7779$, and $V_n^{(cusp)}\approx2.4454$. This point is labeled by an orange circle on Fig.\ref{2LEqPoints}(a), where we may qualitatively observe that the surface does fold after this point.
	
	Similarly, for a given $R_n > R_n^{(cusp)}$, we can calculate the values of $V^*$ and $V_n$ where the dependency $V^*(V_n)$ becomes multi-valued or single-valued corresponding the the saddle-node bifurcation points. With $R_n=8$, these points are calculated as $V^{(1)}\approx 1.4336$ and $V^{(2)}\approx 2.6689$, with corresponding bias voltages $V_n^{(1)}\approx5.4501$ and $V_n^{(2)}\approx2.9232$. These are labeled on Fig.\ref{2LEqPoints}(a) using red circles and agree strongly with the qualitative observations of discontinuities described earlier in this section.
	
	The approaches used for calculating both of these are detailed in Appendix \ref{appendix:cusp}.
	
	\section{Stability of Equilibrium Points and Self-Oscillations}
	
	Next, we investigate linear stability of the revealed equilibria by inspecting the eigenvalues of the Jacobian of the system  (\ref{eq:nmod}), $\mathbf{J}$, calculated at the equilibrium values of $X=0$, $Y=0$, $Z=Z_T$ and $V=V^*$ as stated in Eq. (\ref{eq:eq})
	\begin{equation}
		\begin{gathered}
			\label{eq:Jac}
			\mathbf{J}(0,0,Z_T,V^*) =\\
			\scriptsize
			\begin{pmatrix}
				\frac{\partial\dot{V}}{\partial V}\Bigr|_{V=V^*} & \frac{R_n}{2}(F_{1,1}^*-F_{0,0}^*)V^*&-R_nF_{0,1}^*V^*&0\\0&-\alpha\Gamma&0&0\\2Z_T\frac{\partial\dot{V}}{\partial V}\Bigr|_{V=V^*}&Z_TR_n(F_{1,1}^*-F_{0,0}^*)V^*&-2Z_TR_nF_{0,1}^*V^*-\Gamma &\Omega\\0&0&-\Omega&-\Gamma
			\end{pmatrix},\\
		\end{gathered}
	\end{equation}
	\begin{equation*}
		\begin{gathered}
			\frac{\partial\dot{V}}{\partial V}\Bigr|_{V=V^*}= -\left\{1+\frac{R_n}{2}\left[(1+Z_T)(F_{0,0}^*-l\sqrt{2}F_{0,0}'^* V^*)\right.\right.\\\left.\left.+(1-Z_T)(F_{1,1}^*-l\sqrt{2}F_{1,1}'^*V^*)\right]\right\},
		\end{gathered}
	\end{equation*}
	where $F_{i,j}^* = F_{i,j}(x_0-l\sqrt{2}V^*)$ and $F_{i,j}'^* = dF_{i,j}/dx_V$ at $x_0-l\sqrt{2}V^*$. 
	
	Notably, while the positions of the equilibria in the phase space are determined only by the diagonal elements of the matrix $\left[\sech((\tilde{x}+x_V)/\lambda)\right]$, the Jacobian (\ref{eq:Jac}) depends  also on off-diagonal elements of $\rho$ and $\left[\sech((\tilde{x}+x_V)/\lambda)\right]$. The latter suggests that  quantum coherent processes are involved in stability of the discussed equilibria. 
	The eigenvalues  $\mu$ of the Jacobian could be found by solving the characteristic equation $\det\left[\mathbf{J}-\mu \mathbf{I}\right]=0$. The  solution in analytic form, which could be found e.g. using MATLAB \cite{MATLAB}, is presented separately in  the Appendix \ref{appendix:Eig}.
	
	\begin{figure}
		\centering
		\begin{subfigure}[b]{0.5\textwidth}
			\centering
			\includegraphics[width=\linewidth]{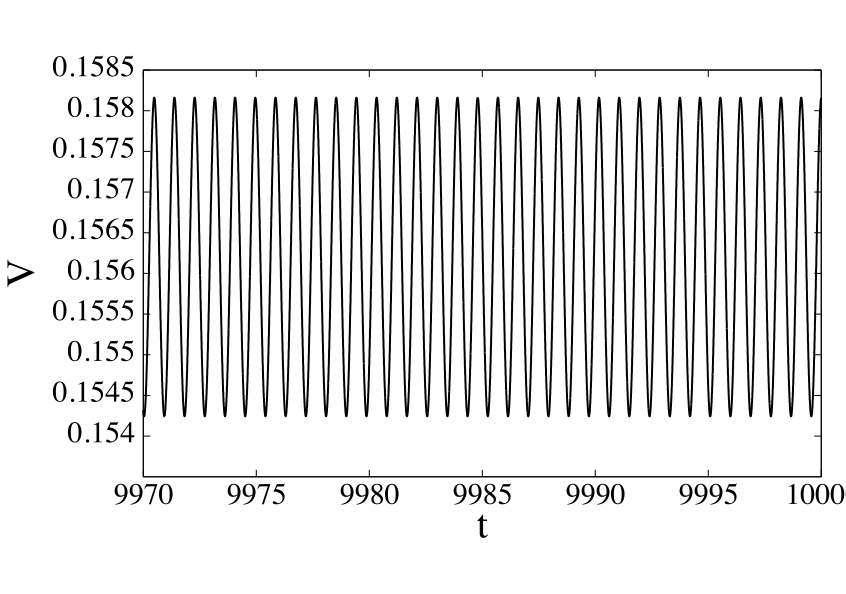}
			\caption{}
			\label{SelfOscillations,A}
		\end{subfigure}
		\begin{subfigure}[b]{0.5\textwidth}
			\centering
			\includegraphics[width=\linewidth]{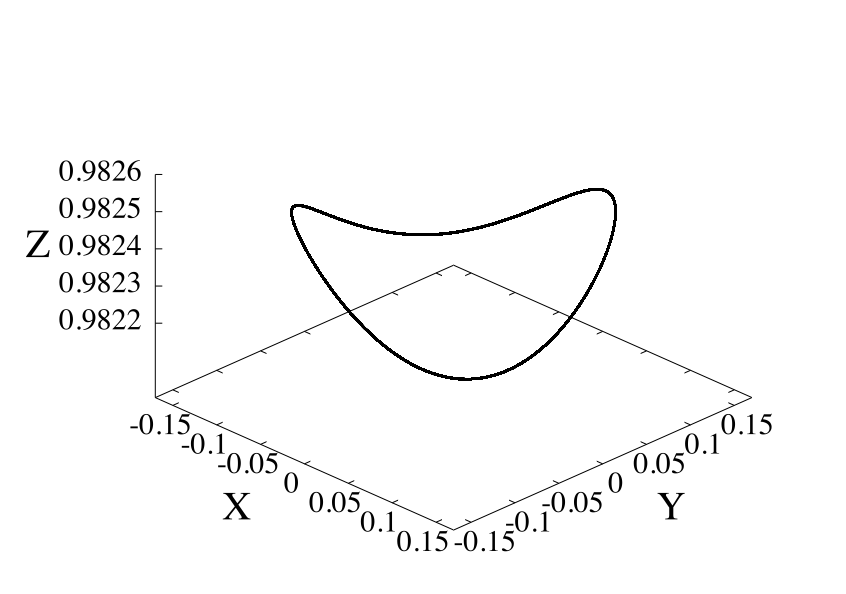}
			\caption{}
			\label{SelfOscillaitons,B}
		\end{subfigure}
		\begin{subfigure}[b]{0.5\textwidth}
			\centering
			\includegraphics[width=\linewidth]{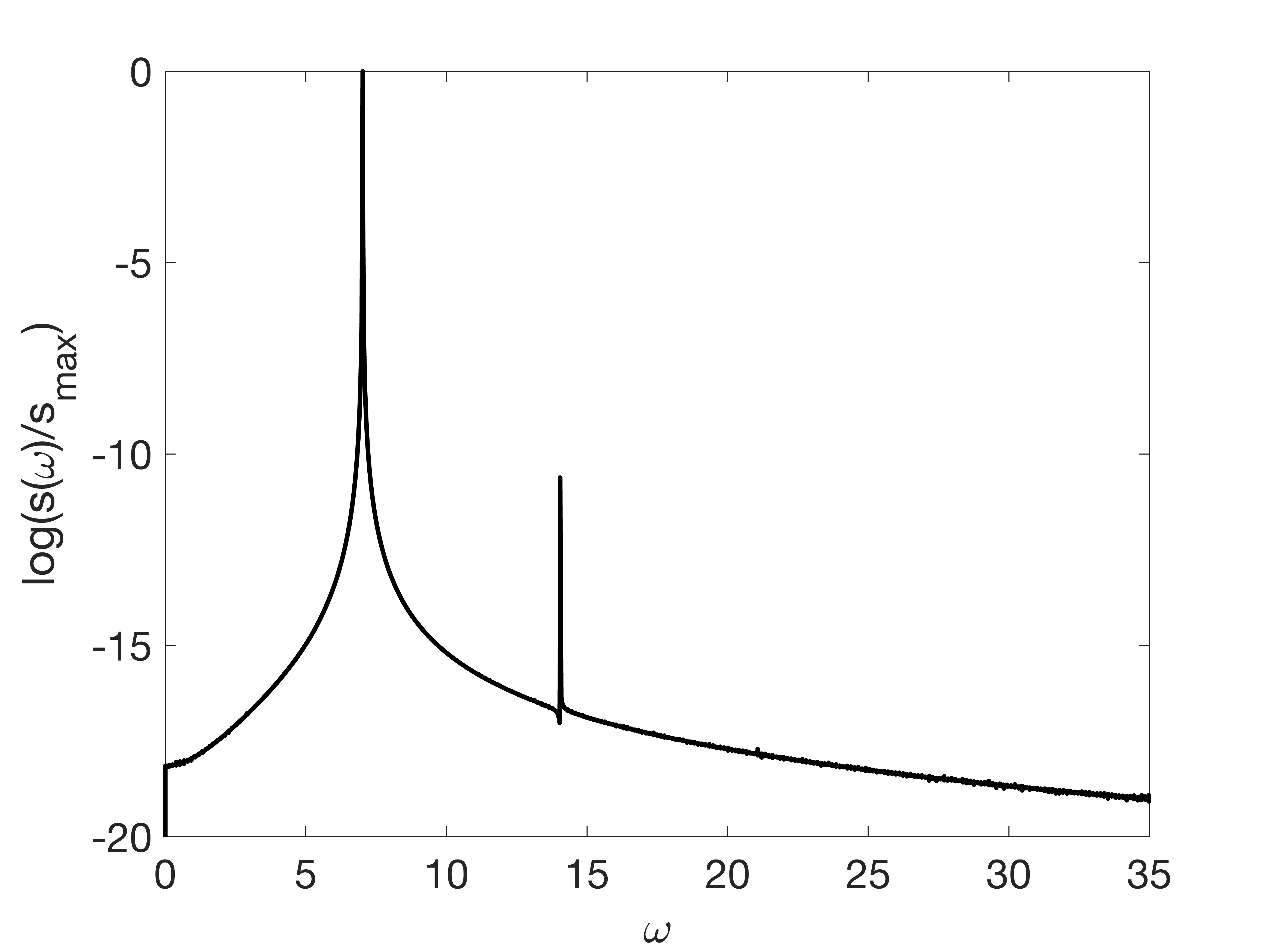}
			\caption{}
			\label{SelfOscillaitons,C}
		\end{subfigure}
		\caption{Self-oscillations observed when $R_n = 5$ $V_n = 0.23$, $\Gamma=0.1$, $\alpha=1$, and $Z_T = 1$. a) Long term voltage dynamics showing spiking behavior. b) Stable limit cycle in Bloch vector components variables $X$, $Y$, and $Z$. c) Power spectrum of oscillations as a function of frequency $\omega$.}
		\label{SelfOscillations}
	\end{figure}
	In addition to the discussed above saddle-node bifurcations characterized by a single zero-valued real eigenvalue, the analysis reveals occurrence of the Andronov-Hopf bifurcations \cite{Anishchenko1995}, which are distinguished by two complex-conjugate eigenvalues with zero-valued real parts. In our system, the onset of  Andronov-Hopf bifurcations  gives birth to stable self-sustained oscillations. These oscillations are robust to the parameters'  change  and exist  within a range of values of $V_n$, $R_n$, and $Z_T$.
	
	Figure  \ref{SelfOscillations} presents a typical time-series of $V(t)$ near to the bifurcation point (a),  the corresponding  projection of the phase trajectory onto the phase space $(X,Y,Z)$ (b) and  power spectral $S(\omega)$ density of $V(t)$ (c). The plots illustrate periodic character of stationary voltage self-oscillations (a), which in the phase space of the system are represented by a closed curve (b). This closed curve is located inside the Bloch sphere, which evidences the effect of relaxation processes in the system. The  spectrum of the oscillations contains multiple harmonics of the main frequency $\omega=7.0246$ reflecting nonlinear mechanism of generation. Notably, this frequency agrees well with the imaginary part of the the complex eigenvalue of the equilibrium point here, $7.0224$, as expected close to an Andronov-Hopf bifurcation.
	
	\begin{figure*}
		\centering
		\begin{subfigure}[b]{0.49\textwidth}
			\centering
			\includegraphics[width=\linewidth]{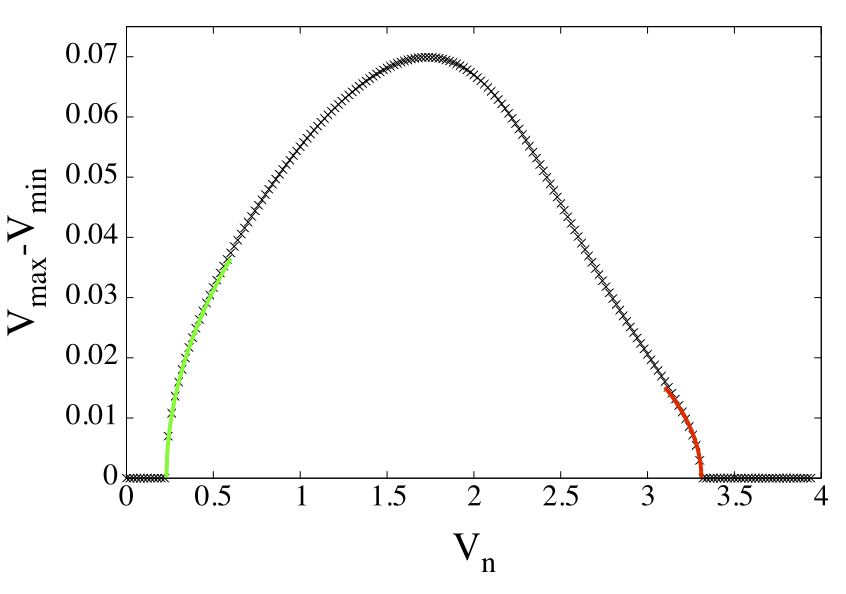}
			\caption{}
			\label{OscVn,A}
		\end{subfigure}
		\begin{subfigure}[b]{0.49\textwidth}
			\centering
			\includegraphics[width=\linewidth]{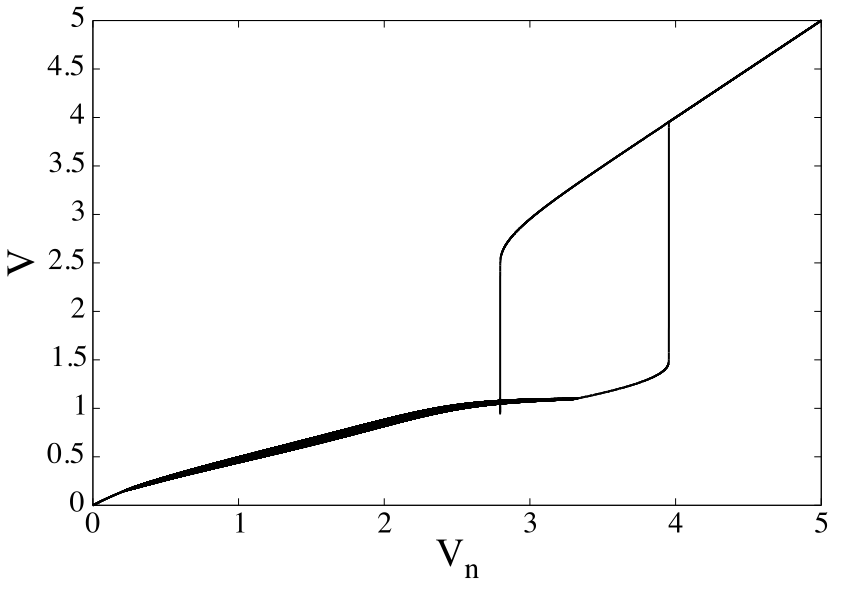}
			\caption{}
			\label{OscVn,B}
		\end{subfigure}
		\caption{ Simulations for  $R_n=5$, $\Gamma=0.1$, and $Z_T=1$ with varying bias voltage $V_n$: (a) span of self-oscillations (black crosses),  fit $0.060\sqrt{V_n-0.23}$ (green curve) for $V_n$ in the range 0.23 to 0.6,  fit $0.033\sqrt{3.31-V_n}$  (red curve)  for $V_n$ in the range 3.1 to 3.31; (b) the dependence of $V(t)$ upon $V_n$.}
		\label{OscVn}
	\end{figure*}
	
	The dependence  of the span of generated self-oscillations $V_{max}-V_{min}$ on the variation of $V_n$ is shown in Fig.\ref{OscVn}(a); here $V_{max}$($V_{min}$) is the maximal (minimal) value of the stationary oscillations $V(t)$. For small input $V_n$ the oscillations do not exist, and a steady state is the only stationary regime of the circuit.  However, above certain critical value of $V_n\approx0.23$, a  non-damped oscillations  of $V$ appear in the system, with the frequency close to $\Omega=7$. Near this critical value $V^{AH}_n$ the amplitude of the self-oscillations grows as $\sqrt{|V_n-V^{AH}_n|}$ highlighting the development of instability related to the Andronov-Hopf bifurcation \cite{Anishchenko1995}. With further growth of $V_n$, the span of self-oscillations reaches its maximum $0.699$ at $V_n=1.73$ and then decreases until the oscillations disappear through another (inverse) Andronov-Hopf bifurcation at $V_n=3.31$. Remarkably, the appearance  of  the Andronov-Hopf bifurcations still preserves the hysteresis in the system allowing coexistence of self-oscillation and and stable equilibrium  (steady state) for the same value of $V_n$, see Fig. \ref{OscVn}(b), where the thickened curve indicates the presence of self-oscillations. In this case the realization of the specific regime depends on the initial conditions.

	\begin{figure*}
		\centering
		\begin{subfigure}[b]{0.49\textwidth}
			\centering
			\includegraphics[width=\linewidth]{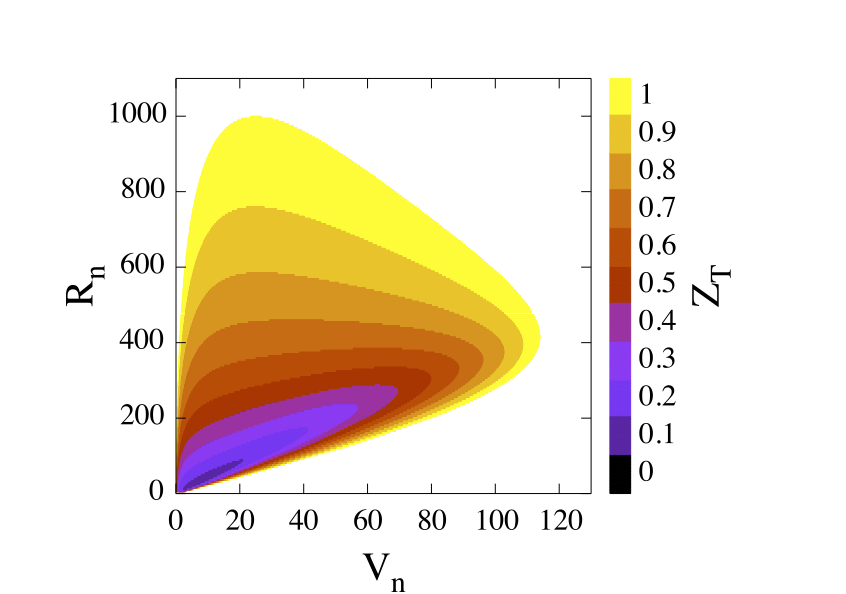}
			\caption{}
			\label{fig:Hopf_region}
		\end{subfigure}
		\begin{subfigure}[b]{0.49\textwidth}
			\centering
			\includegraphics[width=\linewidth]{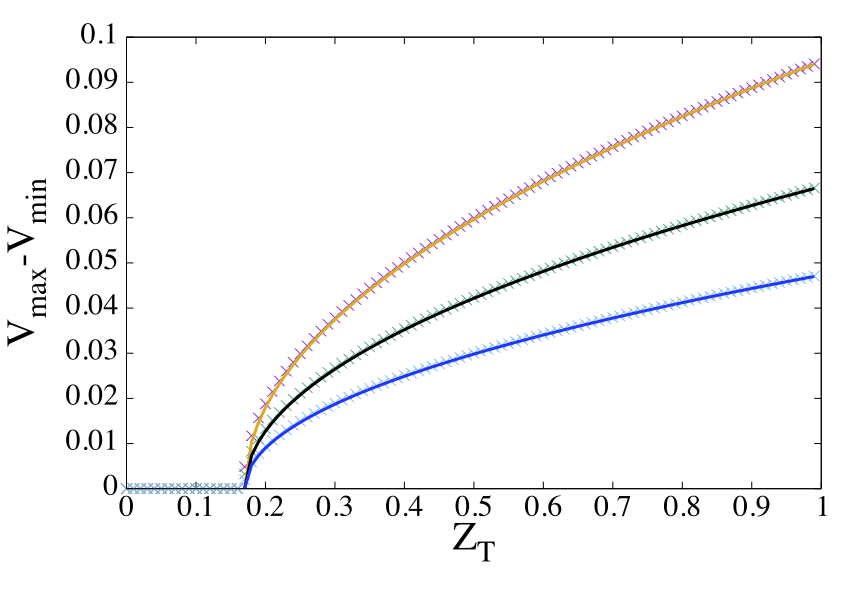}
			\caption{}
			\label{SimulatedOscillations,B}
		\end{subfigure}
		\caption{ (a) The regions of existence of self-sustained oscillations for  $\Gamma=0.1$, $\alpha=1$, different colors reflect different $Z_T$. (b) Points show span of voltage oscillations at different $Z_T$ with $R_n=5$, $V_n=2$, and $\Gamma=0.1$ with (purple) $\alpha=2$, (green) $\alpha=1$, and (blue) $\alpha=0.5$; solid curves show fitted dependencies 
			$0.0734\sqrt{2(Z_T-0.17)}$ (orange), $0.0734\sqrt{Z_T-0.17}$ (black), $0.0734\sqrt{(Z_T-0.17)/2}$ (blue) . }
		\label{RegOsc}
	\end{figure*}
	
	The region of existence for self-sustained oscillations in the parameter space ($V_n$, $R_n$, $Z_T$) is shown in Fig.\ref{RegOsc}(a). In this graph the areas of different color display the regions of self-oscillations on the parameter's plane  ($V_n$, $R_n$) for specific values of $Z_T$ between 0 (infinite temperature) and 1 (zero temperature). The boundary of each colored region corresponds to the onset the Andronov-Hopf bifurcation, and a particular color is assigned to a specific value of $Z_T$. As one can see, the region of the self-oscillations  on the plane  ($V_n$, $R_n$) does not exist for $Z_T = 0$, and remains bounded with $Z_T>0$ . This observation becomes more obvious  in Fig. \ref{RegOsc}(b), where the span of stationary voltage oscillations  is shown in dependence on $Z_T$  and with other parameters being fixed. As the graph shows, for the considered set of the parameters, the oscillations occur for $Z_T>0.16$ after the system has undergone an Andronov-Hopf bifurcation.  With $Z_T\to1$  (the temperature tends to zero) the area of the self-oscillations  in the parameter plane ($V_n$, $R_n$)  grows revealing the key role of quantum coherent dynamics in the development of such a regime.  
	
	To gain an empirical insight into the physical mechanisms for appearance and continuation of self-oscillations, we consider the model (\ref{eq:nmod}) near the bifurcation point with $Z_T=1$. In this case, the oscillations of the state variable $Z$ around $Z_T$ become negligibly small, see Fig. \ref{SelfOscillations}(b). Then after differentiating the second equation in (\ref{eq:nmod}) and substituting the corresponding expressions for $\dot{Y}$ and $\ddot{V}$, the initial system approximately reduces to:
	\begin{eqnarray}
		\label{eq:so1}
		\ddot{X} &+& (2\Gamma+2R_nF_{0,1}(x_V)V)\dot{X}+\Tilde{\Omega}^2X=\epsilon(X, V),\\
		\label{eq:so2}
		\dot{V}  &=&  V_n - \left(1+R_n\left[F_{0,0}(x_V)+ XF_{0,1}(x_V)\right]\right)V,  
	\end{eqnarray}
	where $\Tilde{\Omega}^2=\Omega^2+\Gamma^2$.  The equation (\ref{eq:so1}) allows interpretation of our model as a pendulum with non-trivial state-dependent friction and eigenfrequency $\Tilde{\Omega}$, where (\ref{eq:so2}) provides an additional nonlinearity through the function $\epsilon(X, V)$. This approximation has a form typical for a generic self-oscillator \cite{AKW66}. Actually, the system involves a ``resonator" described by the simple harmonic potential and characterized by the eigenfrequency $\Omega$. It has  energy dissipation defined by the parameters $\Gamma$ and $R_n$, and is supplied with a mechanism for energy pumping into the resonator. Since $F_{0,1}(x_V)$ can have negative values depending on $x_V$,  it is capable to provide the system with negative friction, thus compensating energy loss and  maintaining the oscillation. The representation (\ref{eq:so1}), (\ref{eq:so2}) suggests that for small oscillations (e.g. when the system is close to the Andronov-Hopf bifurcation), the frequency of the oscillations should be close to $\Tilde{\Omega}>\Omega$, which is shown by both spectral analysis of the numerical simulation and calculation of the imaginary part of  the eigenvalues for the equilibria, which are both greater than $\Omega$ near the bifurcation point, see e.g. Fig.\ref{SelfOscillations} and the related discussion above.  
	
	Remarkably, the negative friction relates to the off-diagonal elements of density matrix $\rho\left[\sech((\tilde{x}+x_V)/\lambda)\right]$, confirming the key role of quantum coherent processes in the mechanism of the self-oscillations generation. The latter explains the observation that for  temperature $T\to\infty$ ($Z_T\to0$), the self-oscillations become impossible (see Fig.\ref{RegOsc}(a) and \ref{RegOsc}(b)), since growth of the temperature reduces the coherence of the system and violates the compensation of the energy loss.
	
\begin{figure}
		\centering
		\begin{subfigure}[b]{0.5\textwidth}
			\centering   
			\includegraphics[width=\linewidth]{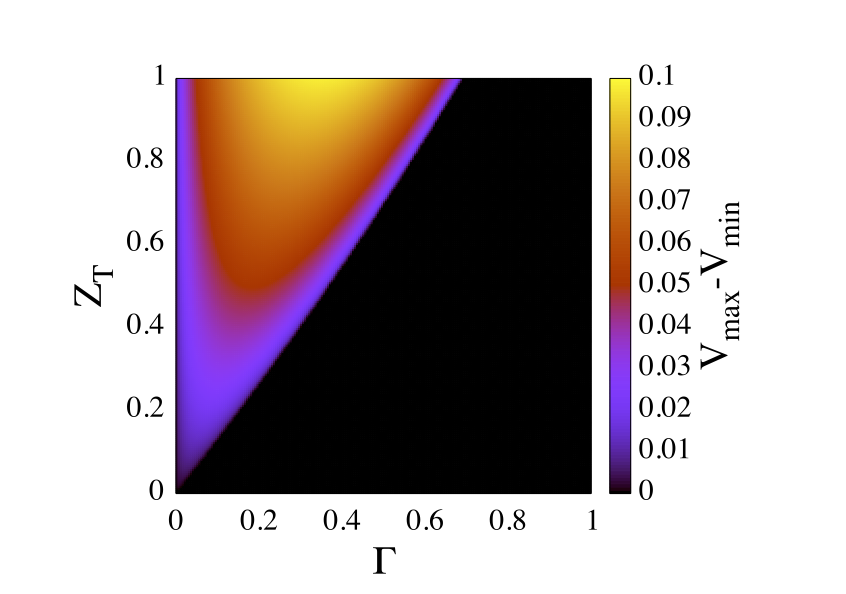}
			\caption{}
			\label{SimulatedOscillations,A}
		\end{subfigure}
		\begin{subfigure}[b]{0.5\textwidth}
			\centering
			\includegraphics[width=\linewidth]{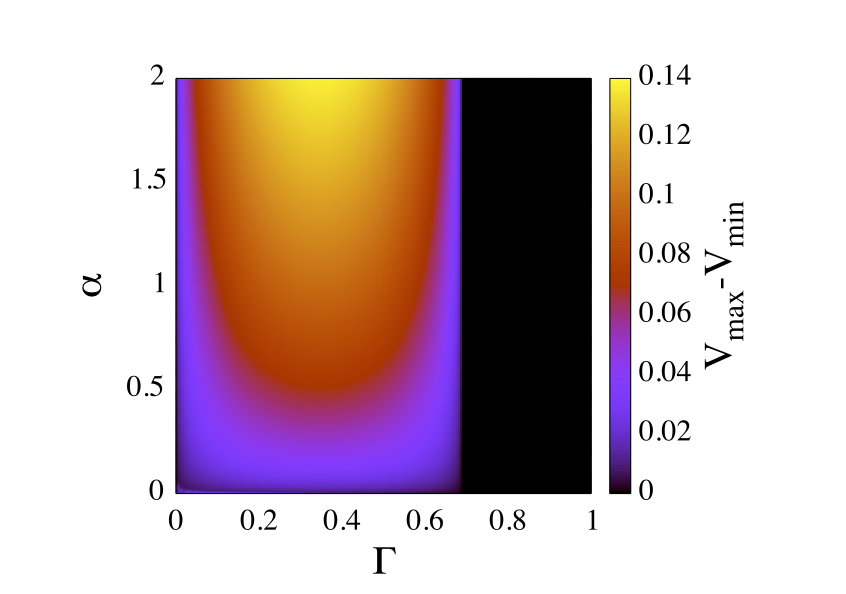}
			\caption{}
			\label{SimulatedOscillations,C}
		\end{subfigure}
		\caption{Heat maps showing how the span of long term oscillations in $V$ depend on dissipation parameters of the quantum system with fixed external voltage, $V_n = 2$, and external resistance, $R_n = 6$. (a) Heat map showing $V_{max}-V_{min}$ when $\alpha = 1$ as $\Gamma$ and $Z_T$ are changed. (b) Heat map showing $V_{max}-V_{min}$ when $Z_T = 1$ as $\Gamma$ and $\alpha$ are changed.}
		\label{cont}
	\end{figure}

	Finally, we consider how the relaxation parameter $\Gamma$ and the ratio between the relaxation rate and pure dephasing rate $\alpha$ affect the self-oscillations. Fig \ref{cont}(a) illustrates the span of the oscillations (represented by color) as a function of the relaxation parameters $\Gamma$ and $Z_T$ for other parameters fixed as $R_n = 6$, $V_n = 2$, and $\alpha=1$. With increase of $Z_T$ the area of the oscillations' existence grows, which reflects the fact that the coherence of the system important for energy pumping becomes more prominent at lower $T$.  The stronger coherence explains also the growth of the maximal amplitude of the self-oscillations as $Z_T\to1$. On the other hand,  there is always a maximal relaxation rate in $\Gamma$, above which the pumping energy rate is not enough to compensate the rate of energy loss. These arguments explain the tongue-like shape of the oscillations' existence region.  Fig \ref{cont}(b) reveals that this maximal value of  $\Gamma$ bounding the self-oscillations does not depend on the ratio $\alpha$.
	
	Interestingly, the amplitude of self-oscillations depends both on $\Gamma$ and $\alpha$ featuring a possibility for the optimal ratio between the energy dissipation, dephasing and energy pumping rates to maximize the amplitude the generated  oscillations.  Determining the conditions under which such optimization could be achieved is beyond the scope of this work and requires further investigation.
	
	\section{Conclusion} 
	In conclusion, we propose a model for a simple quantum coherent element with memristive properties, whose physical operational principle is based on the charge tunneling between two terminals through a quantum particle shuttling within a simple harmonic potential.  We show that such a device is characterized by the conductance which non-monotonically depends on the voltage drop between the terminals. Being included as an active element in an electric circuit of an artificial neuron it enables the hysteretic input-output characteristics, which can be  utilized in the memory components of neuromorphic systems. 
	
	In addition, within the range of the parameters  the circuit is able to demonstrate self-sustained oscillations mimicking the behavior of spiking neurons.  Moreover, the appearance of the self-oscillations does not destroy the hysteresis in the system thus allowing the coexistence of the steady state and self-oscillations for the same parameters of the circuit. The study shows that both hysteretic  and self-oscillating behavior are robust phenomena existing in wide range of circuit parameters and the parameters of the quantum device.
	Thus, utilization of the proposed quantum element endows the artificial neuron with a powerful potential for controlling and combining  steady states, self-oscillations and hysteretic regimes by simple tuning the circuit parameters or/and applied voltage bias.
	
	Applying a  combination of analytical and numerical methods we investigated the origin and the mechanisms for the hysteretic and  self-oscillation phenomena and disclosed the instabilities leading to their emergence. In particular our analysis revealed that the stability properties of the circuit are dependent on the coherence of the quantum element; while the hysteretic behavior could exist even without influence of the quantum coherent processes, self-oscillations require their presence. The reason for this is quantum coherent processes allow a coherent pumping of energy to compensate losses due to relaxation and dephasing. Such an interplay of the pumping and dissipation provides conditions required for the development of the instability associating with the Andronov-Hopf bifurcation, which gives rise to self-sustained current and voltage oscillations in the circuit.
	
	The reported findings shed new light on the  principles for designing memristive quantum systems and on quantum regimes which could be utilized for their control. They also provide an insight on the use of the quantum switchers with memory in classical circuits and suggest a way to recognize the existence of quantum coherent processes in the dynamics of a classical circuit featuring a quantum memristive device.   Thus, our results inform future strategies for further development of quantum memristive elements and their applications to the neuromorphic circuits.
	
	\section*{Acknowledgements}
	We acknowledge the support of the UK Research and Innovation (UKRI) Horizon Europe guarantee scheme for the QRC-4-ESP project under grant number 101129663.
	
	\appendix
	\section{Derivation of Hamiltonian}
	\label{appendix:Hamiltonian}
	Here we derive the form of the Hamiltonian we will use throughout this paper from our initial Hamiltonian stated in Eq.(\ref{eq:H}).
	
	From our initial Hamiltonian,
	\begin{gather*}
		\mathcal{H} = -\frac{\hbar^2}{2m}\frac{\partial^2}{\partial x^2} +\frac{1}{2}m\Omega_p^2(x-x_0)^2 +\frac{qV_mx}{2L},
	\end{gather*}
	we can collate the perturbation into the quadratic term by completing the square as
	\begin{equation}
		\begin{gathered}
			\mathcal{H} = -\frac{\hbar^2}{2m}\frac{\partial^2}{\partial x^2} +\frac{1}{2}m\Omega_p^2\left[(x-x_0)^2 + \frac{qV_mx}{m\Omega_p^2L}\right]
			\\=-\frac{\hbar^2}{2m}\frac{\partial^2}{\partial x^2} +\frac{1}{2}m\Omega_p^2\left\{\left(x-x_0+\frac{qV_m}{2m\Omega_p^2L}\right)^2\right.\\\left.+\frac{1}{2}m\Omega_p^2\left[x_0^2-\left(\frac{qV_m}{2m\Omega_p^2L}-x_0\right)^2\right]\right\}.
		\end{gathered}
	\end{equation}
	If we assume that $V_m$ is a function of time only then the final term $m\Omega_p^2[x_0^2-(qV_m/2m\Omega_p^2L-x_0)^2]/2$ is also a function of time only and therefore can be removed by use of an appropriate phase factor leaving the Hamiltonian as
	\begin{gather*}
		\mathcal{H} = -\frac{\hbar^2}{2m}\frac{\partial^2}{\partial x^2} +\frac{1}{2}m\Omega_p^2\left(x-x_0+\frac{qV_m}{2m\Omega_p^2L}\right)^2
	\end{gather*}
	We may then make the transformation of variables $\tau=t$, $\tilde{x} = x-x_0+qV_m/2m\Omega_p^2L$, leading to a change in partial derivatives:
	\begin{align*}
		\frac{\partial}{\partial x}&\rightarrow\frac{\partial\Tilde{x}}{\partial x}\frac{\partial}{\partial\Tilde{x}}+\frac{\partial\tau}{\partial x}\frac{\partial}{\partial\tau} = \frac{\partial}{\partial\Tilde{x}}\\
		\frac{\partial}{\partial t}&\rightarrow\frac{\partial\Tilde{x}}{\partial t}\frac{\partial}{\partial\Tilde{x}}+\frac{\partial\tau}{\partial t}\frac{\partial}{\partial\tau} = \frac{q}{2m\Omega_p^2L}\frac{dV_m}{dt}\frac{\partial}{\partial\Tilde{x}}+\frac{\partial}{\partial\tau}.
	\end{align*}
	Considering then Schr\"{o}dingers equation; in our initial variables it should take the form
	\begin{gather*}
		i\hbar\frac{\partial}{\partial t}\Psi = \left(-\frac{\hbar^2}{2m}\frac{\partial^2}{\partial x^2} +\frac{1}{2}m\Omega_p^2(x-x_0+\frac{qV_m}{2m\Omega_p^2L})^2\right)\Psi
	\end{gather*}
	Which transforms with the new variables to
	\begin{equation}
		\begin{gathered}
			i\hbar\left(\frac{q}{2m\Omega_p^2L}\frac{dV_m}{dt}\frac{\partial}{\partial\Tilde{x}}+\frac{\partial}{\partial\tau}\right)\Psi = \left(-\frac{\hbar^2}{2m}\frac{\partial^2}{\partial\Tilde{x}^2} + \frac{1}{2}m\Omega_p^2\Tilde{x}^2\right)\Psi\\\implies\\
			i\hbar\frac{\partial}{\partial\tau}\Psi = \left(-\frac{\hbar^2}{2m}\frac{\partial^2}{\partial\Tilde{x}^2} + \frac{1}{2}m\Omega_p^2\Tilde{x}^2 - i\hbar\frac{q}{2m\Omega_p^2L}\frac{dV_m}{dt}\frac{\partial}{\partial\Tilde{x}}\right)\Psi.
		\end{gathered}
	\end{equation}
	Therefore, we should take the Hamiltonian as
	\begin{gather*}
		\mathcal{H} = -\frac{\hbar^2}{2m}\frac{\partial^2}{\partial\Tilde{x}^2} + \frac{1}{2}m\Omega_p^2\Tilde{x}^2 - i\hbar\frac{q}{2m\Omega_p^2L}\frac{dV_m}{dt}\frac{\partial}{\partial\Tilde{x}}.
	\end{gather*}
	
	\section{Thermal Parameter Derivation}
	\label{appendix:Therm}
	In this appendix we intend to show that $Z_T$ can be characterized by the temperature of the system.
	
	Assuming $\gamma_T$ and $\gamma_\varphi$ are nonzero and do not depend on any external circuit parameters (e.g. $R_{ext}$ or $V_{ext}$ in the artificial neuron circuit shown in Fig.\ref{DeviceAndCircuit}(a)),  without loss of generality, we can consider the circuit such that $dV_m/dt = 0$ (i.e. $R_{ext} = 0$ and $V_{ext} = 0$ with $V_m(0)=0$ for the artificial neuron circuit equation given in Eq.(\ref{eq:V})). In this case the system (\ref{eq:xyz}) takes the form:
	\begin{equation}
		\begin{gathered}
			\frac{dX}{dt} = \Omega_p Y - \gamma_\varphi X\\
			\frac{dY}{dt} = -\Omega_p X - \gamma_\varphi Y\\
			\frac{dZ}{dt} = -\gamma_T (Z-Z_T)
		\end{gathered}
	\end{equation}
	Which can be solved exactly as
	\begin{equation}
		\label{exactSol}
		\begin{gathered}
			Z(t) = Z_T + (Z_0-Z_T)e^{-\gamma_T t}\\
			X(t) = A \sin(\Omega_p t+\phi)e^{-\gamma_\varphi t}\\
			Y(t) = A \cos(\Omega_p t+\phi)e^{-\gamma_\varphi t}.
		\end{gathered}
	\end{equation}
	Where $A^2+Z_0^2=1$ is a requirement for beginning in a pure quantum state. This will clearly converge to the thermal state $X=Y=0$, $Z=Z_T$.\\
	Now we may consider the probability of being found in either state as being given by both the diagonal elements of the density matrix and a canonical ensemble depending on the temperature at equilibrium:
	\begin{gather}
		\zeta_0 = \frac{1}{2}(1+Z_T) = \frac{1}{\xi}e^{-\frac{\hbar\Omega_p}{2k_b T}}\label{zeta0}\\
		\zeta_1 = \frac{1}{2}(1-Z_T) = \frac{1}{\xi}e^{-\frac{3\hbar\Omega_p}{2k_b T}}\label{zeta1}\\
		\xi = e^{-\frac{3\hbar\Omega_p}{2k_b T}}+e^{-\frac{\hbar\Omega_p}{2k_b T}}
	\end{gather}
	Where $\zeta_i$ denotes the probability of the system being in the $i^{th}$ energy level, given by the diagonal elements of the density matrix.\\
	By subtracting Eq.(\ref{zeta1}) from Eq.(\ref{zeta0}), we can isolate $Z_T$ and therefore show that
	\begin{equation}
		Z_T = \tanh{\frac{\hbar\Omega_p}{2k_b  T}}
	\end{equation}
	This result can also be found in literature, such as in \cite{Zagoskin2011}. This gives a physical interpretation of $Z_T$ as describing the temperature of the system. It also bounds $Z_T$ as $Z_T\in[0,1]$ corresponding to the infinite and zero temperature limits respectively.
	
	\section{Constraints on Dephasing Parameters}
	\label{appendix:Dephasing}
	In this appendix we intend to justify the necessary constraint that the ratio $\gamma_T/\gamma_\varphi<2$.
	
	Under the same assumptions as Appendix \ref{appendix:Therm}, we can again consider the case where $dV_m/dt = 0$ and obtain the solution for $X$, $Y$, and $Z$ given in Eq.(\ref{exactSol}). 
	We can now take the trace of $\rho^2$, which, from the positivity of probability, is bounded between $\frac{1}{2}$ and $1$ \cite{Joos2013}, and substitute these solutions into this to give
	\begin{equation}
		\begin{gathered}
			\Tr{\rho^2} = \frac{1}{2}(1+Z_T^2 +2(Z_0-Z_T)Z_Te^{-\gamma_T t} \\+ (Z_0-Z_T)^2e^{-2\gamma_T t}+ A^2e^{-2\gamma_\varphi t})
		\end{gathered}
	\end{equation}
	As we have previously shown $Z_T$ is a function of temperature only, we can choose to consider when $Z_T = 1$ in the zero temperature limit (with the implicit assumption that $\gamma_T$ and $\gamma_\varphi$ also do not depend on $Z_T$).\\
	Comparing exponents, we can see that $e^{-2\gamma_T t}$ will decay much faster than $e^{-\gamma_T t}$ and so we may neglect this term leaving
	\begin{gather*}
		\Tr{\rho^2} = \frac{1}{2}(2 +2(Z_0-1)e^{-\gamma_T t}+ A^2e^{-2\gamma_\varphi t})
	\end{gather*}
	If we consider $2\gamma_\varphi<\gamma_T$ then it is clear that the $e^{-\gamma_T t}$ will at some point become negligible in comparison with the $e^{-2\gamma_\varphi t}$ term leaving
	\begin{equation*}
		\Tr{\rho^2} = \frac{1}{2}(2+ A^2e^{-2\gamma_\varphi t})
	\end{equation*}
	This will, for a general initial conditions, be greater than $1$ for all $t>0$ which corresponds to the Bloch vector evolving outside of the Bloch sphere with nonphysical probabilities.\\
	If we instead impose $2\gamma_\varphi>\gamma_T$ then we will eventually have 
	\begin{equation*}
		\Tr{\rho^2} = \frac{1}{2}(2+2(Z_0 - 1)e^{-\gamma_T t})
	\end{equation*}
	Where $Z_0 - 1 \in [-2,0]$ resulting in evolution which is always in the Bloch sphere. As such, we require that the ratio $\gamma_T/\gamma_\varphi < 2$, which can be built into the system by the parameter transformations $\gamma_\varphi = \gamma$ and $\gamma_T = \alpha\gamma$ where $\alpha\in[0,2]$. This result is in accordance with the conditions of a two-dimensional Lindblad equation established in \cite{Bellac2006, Zagoskin2011}.

	\section{$F_{i,j}(x_V) functions$}
	\label{Appendix:F}
	\begin{figure}[h]
		\centering
		\includegraphics[width=0.9\linewidth]{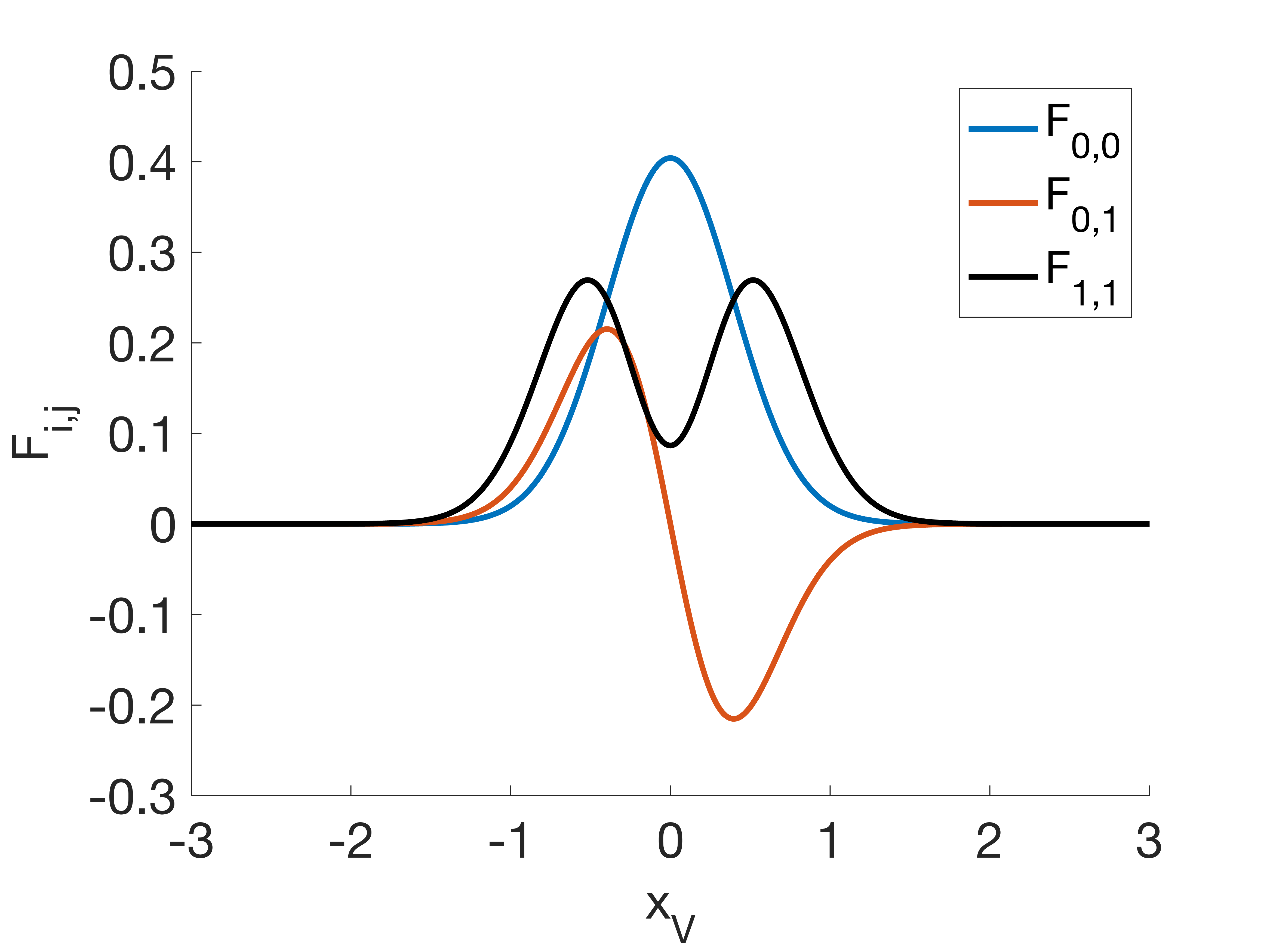}
		\caption{Plots of $F_{i,j}(x_V)$ with $l=0.5$ and $\lambda=0.13$.}
		\label{FijPlots}
	\end{figure}
	
	\section{Bifurcation analysis using $f(V)$}
	\label{appendix:cusp}
	In this appendix we present analysis of $f(V)$, defined in Eq.\ref{eq:f(V)}, to predict the values $R_n^{(cusp)}$, $V_n^{(cusp)}$, and $V^{(cusp)}$ where the cusp bifurcation occurs as well as the values of $V_n^{(1,2)}$ and $V^{(1,2)}$ for a given $R_n > R_n^{(cusp)}$ where the we expect saddle-node bifurcations defining the boundaries of hysteresis for that choice of $R_n$. 
	
	To find the critical value of $R^{(cusp)}_n$ above which the hysteresis appears, we consider the conditions under which the equation $f(V)$ in (\ref{eq:f(V)}) may have three roots for a particular $V_n$. Such situation requires $f(V)$ to have both a maximum and a minimum. The latter means that the function
	\begin{equation*}
		\begin{gathered}
			D_V f(V) = -1 - \frac{R_n}{2}\{(1+Z_T)[F_{0,0}(x_0-l\sqrt{2}V)\\-l\sqrt{2}F_{0,0}'(x_0-l\sqrt{2}V)V]
			\\+(1-Z_T)[F_{1,1}(x_0-l\sqrt{2}V)\\-l\sqrt{2}F_{1,1}'(x_0-l\sqrt{2}V)V]\}
		\end{gathered}
	\end{equation*}
	should have a two roots. Note that $D_V$ denotes $d/dV$ and $'$ denotes $d/dx_V$. Given the definition of $F_{i,j}$, it is clear that as $|V|\rightarrow\infty$ both $F_{i,j}$ and $F_{i,j}'$ will tend to $0$ so $D_Vf(V)\rightarrow -1$. From this we can infer that two roots will require an $D_Vf(V)$ to have an extremum to be a local maximum greater than greater than $0$. Therefore the transition between cases when the equation $f(V)$ has three roots and  one root requires that (i)  $D_V^2 f(V)=0$ ensuring the existence of the maximum for $D_V f$ and simultaneously (ii)  $D_V f=0$, since in this case the maxima and minima of $f(V)$ collide. The condition (ii) yields 
	\begin{equation}
		\begin{gathered}
			\label{eq:cuspV}
			(1+Z_T)(lF_{0,0}''(x_0-l\sqrt{2}V^*)V^*-\sqrt{2}F_{0,0}'(x_0-l\sqrt{2}V^*))+\\(1-Z_T)(lF_{1,1}''(x_0-l\sqrt{2}V^*)V^*-\sqrt{2}F_{1,1}'(x_0-l\sqrt{2}V^*)) \\= 0
		\end{gathered}
	\end{equation}
	returning the critical value $V=V^{(cusp)}$ corresponding to the cusp. This equation will have multiple solutions, only those corresponding to local maxima of $D_Vf(V)$ are voltages where a cusp bifurcation will occur, that is $V$ such that $D_V^2f(V)=0$ and $D_V^3f(V)<0$ (graphical analysis of Eq. \ref{eq:cuspV} is likely sufficient to determine which roots correspond to local maxima). Of these solutions, there will be two corresponding to local maxima of $D_Vf(V)$ demonstrating the two separate regions of hysteresis as shown in Fig.\ref{2LEqPoints}(c).
	Substituting $V^{(cusp)}$ into the equation $D_V f(V^{(cusp)})=0$ and rearranging for $R_n$ yields
	\begin{equation}
		\begin{gathered}
			\label{eq:cusp1}
			R_n^{(cusp)} = \\-2\cdot\left\{ (1+Z_T)\left[F_{0,0}(x_V^{(cusp)})-l\sqrt{2}F_{0,0}'(x_V^{(cusp)})V^{(cusp)}\right] \right. \\ \left.+(1-Z_T)\left[F_{1,1}(x_V^{(cusp)})-l\sqrt{2}F_{1,1}'(x_V^{(cusp)})V^{(cusp)}\right] \right\}^{-1},
		\end{gathered}
	\end{equation}
	where $ x_V^{(cusp)} = x_0 - l\sqrt{2}V^{(cusp)}$. 
	Then the critical value of the bias $V_n^{(cusp)}$ can be calculated from the original requirement $f(V^{(cusp)})=0$ and rearranging for $V_n$ to give
	\begin{equation}
		\begin{gathered}
			\label{eq:cusp2}
			V_n^{(cusp)} = \left\{1+\frac{R_n^{(cusp)}}{2}\left[(1+Z_T)F_{0,0}(x_0-l\sqrt{2}V^{(cusp)})\right.\right.\\\left.\left.+(1-Z_T)F_{1,1}(x_0-l\sqrt{2}V^{(cusp)})\right]\right\}V^{(cusp)}
		\end{gathered}
	\end{equation}
	
	We note that from the viewpoint of the bifurcation theory, the Eqs. (\ref{eq:cusp1}) and  (\ref{eq:cusp2}) imply the conditions for emergence of the cusp bifurcation (or cusp catastrophe) \cite{SStrogatz} in our system.
	
	For $R_n>R_n^{(cusp)}$ the condition for saddle-node bifurcations (orange circles in Fig. \ref{2LEqPoints}(b)) limiting the area of hysteresis existence, are defined by the extrema of $f(V)$  at  $V^{(1,2)}$. The latter can be found from the equation
	
	\begin{equation}
		\begin{gathered}
			\label{eq:s-n}
			D_V f(V^{(i)}) = -1-\frac{R_n}{2}\left\{(1+Z_T)\left[F_{0,0}(x_0-l\sqrt{2}V^{(i)})
			-\right.\right.\\\left.l\sqrt{2}F_{0,0}'(x_0-l\sqrt{2}V^{(i)})V^{(i)}\right]+\\
			(1-Z_T)\left[F_{1,1}(x_0-l\sqrt{2}V^{(i)})
			-\right.\\\left.\left.l\sqrt{2}F_{1,1}'(x_0-l\sqrt{2}V^{(i)})V^{(i)}\right]\right\}=0.
		\end{gathered}
	\end{equation}
	
	Substituting  $V^{(1,2)}$ in the equation $f(V^{(1,2)})=0$ we can once again rearrange to give the bias voltage(s) $V_n$ where the saddle-node bifurcation occurs:
	\begin{equation}
		\begin{gathered}
			\label{eq:s-n2}
			V_n^{(i)} = \left\{1+\frac{R_n}{2}\left[(1+Z_T)F_{0,0}(x_0-l\sqrt{2}V^{(i)})\right.\right.\\\left.\left.+(1-Z_T)F_{1,1}(x_0-l\sqrt{2}V^{(i)})\right]\right\}V^{(i)},~ i=1,2. 
		\end{gathered}
	\end{equation}
	For  $R_n=8$, Eqs. \ref{eq:s-n} and \ref{eq:s-n2} yields $V^{(1)}\approx 1.4336$ and $V^{(2)}\approx 2.6689$, with corresponding bias voltages $V_n^{(1)}\approx5.4501$ and $V_n^{(2)}\approx2.9232$.
	
	Note, the similar analysis can be performed to characterize the hysteresis for the negative values of $V_n$, see Fig.\ref{2LEqPoints}(c).
	
	\newpage
	\section{Eigenvalues of Jacobian}
	\label{appendix:Eig}
	\begin{widetext}
		\begin{equation*}
			\begin{gathered}
				\mathbf{Eig}(\mathbf{J}(0,0,Z_T,V^*)) =
				\begin{pmatrix}
					-\Gamma \,\alpha \\
					\frac{\sigma_5 }{6}-\frac{2\,\Gamma }{3}-\frac{F_{0,0}^* \,R_n }{6}-\frac{F_{1,1}^* \,R_n }{6}-\frac{6\,\sigma_6 }{\sigma_5 }-\sigma_3 +\sigma_2 -\sigma_1 +\frac{\sigma_{33} }{6}+\frac{\sigma_{32} }{6}+\frac{\sigma_{31} }{6}-\frac{\sigma_{30} }{6}-\frac{1}{3}\\
					\frac{3\,\sigma_6 }{\sigma_5 }-\frac{\sigma_5 }{12}-\sigma_4 -\frac{F_{0,0}^* \,R_n }{6}-\frac{F_{1,1}^* \,R_n }{6}-\frac{2\,\Gamma }{3}-\sigma_3 +\sigma_2 -\sigma_1 +\frac{\sigma_{33} }{6}+\frac{\sigma_{32} }{6}+\frac{\sigma_{31} }{6}-\frac{\sigma_{30} }{6}-\frac{1}{3}\\
					\frac{3\,\sigma_6 }{\sigma_5 }-\frac{\sigma_5 }{12}+\sigma_4 -\frac{F_{0,0}^* \,R_n }{6}-\frac{F_{1,1}^* \,R_n }{6}-\frac{2\,\Gamma }{3}-\sigma_3 +\sigma_2 -\sigma_1 +\frac{\sigma_{33} }{6}+\frac{\sigma_{32} }{6}+\frac{\sigma_{31} }{6}-\frac{\sigma_{30} }{6}-\frac{1}{3}
				\end{pmatrix}
			\end{gathered}
		\end{equation*}
	\end{widetext}
	where
	\begin{gather*}
		\sigma_1 =\frac{2F_{0,1}^* R_n V^*Z_T }{3}\\
		\sigma_2 =\frac{F_{1,1}^* R_n Z_T }{6}\\
		\sigma_3 =\frac{F_{0,0}^* R_n Z_T }{6}\\
		\sigma_4 =\sqrt{3}(\frac{\sigma_5}{12}+\frac{3\,\sigma_6}{\sigma_5})i\\
		\sigma_5 = \biggl(\sigma_7-\sigma_8+\sigma_9-\sigma_{10}+\sigma_{11}+\sigma_{12} +\sigma_{13}+\sigma_{14}+\sigma_{15}\\+\sigma_{16}-\sigma_{17}+\sigma_{18}-\sigma_{19}-\sigma_{20}-\sigma_{21}-\sigma_{22}-\sigma_{23}\\-\sigma_{24}^3-108(\Gamma^2+\Omega^2)+\left\{\left[24\Gamma+12\Gamma^2-\sigma_{24}^2+12\Omega^2\right.\right.\\+12\Gamma R_n(F_{0,0}^*+F_{1,1}^*)+12\Gamma R_n Z_T(F_{0,0}^*-F_{1,1}^*\\+2F_{0,1}^*V^*)-12\sqrt{2}\Gamma R_n V^* l(F_{0,0}'^*+F_{1,1}'^*+F_{0,0}'^*Z_T\\\left.-F_{1,1}'^*Z_T)\right]^3+\left[\sigma_7-\sigma_8+\sigma_9-\sigma_{10}+\sigma_{11}+\sigma_{12} \right.\\+\sigma_{13}+\sigma_{14}+\sigma_{15}+\sigma_{16}-\sigma_{17}+\sigma_{18}-\sigma_{19}-\sigma_{20}\\\left.\left.\left.-\sigma_{21}-\sigma_{22}-\sigma_{23}-\sigma_{24}^3-108(\Gamma^2+\Omega^2)\right]^2\right\}^{\frac{1}{2}}\right)^{\frac{1}{3}}
		\\
		\sigma_6 =\frac{2\Gamma }{3}+\frac{\Gamma^2 }{3}-\frac{{\sigma_{24} }^2 }{36}+\frac{\Omega^2 }{3}+\frac{F_{0,0}^* \Gamma R_n }{3}+\frac{F_{1,1}^* \Gamma R_n }{3}\\+\frac{F_{0,0}^* \Gamma R_n Z_T }{3}-\frac{F_{1,1}^* \Gamma R_n Z_T }{3}+\frac{\sigma_{29} }{3}-\frac{\sigma_{28} }{3}-\frac{\sigma_{27} }{3}\\-\frac{\sigma_{26} }{3}+\frac{\sigma_{25} }{3}\\
		\sigma_7 =18\sigma_{24} [(2\Gamma +\Gamma^2 +\Omega^2 +F_{0,0}^* \Gamma R_n +F_{1,1}^* \Gamma R_n\\+F_{0,0}^* \Gamma R_n Z_T -F_{1,1}^* \Gamma R_n Z_T +\sigma_{29} -\sigma_{28} -\sigma_{27} -\sigma_{26} +\sigma_{25})]\\
		\sigma_8 =54\sqrt{2}F_{1,1}'^* R_n V^*Z_T l\Omega^2 \\
		\sigma_9 =54\sqrt{2}F_{0,0}'^* R_n V^*Z_T l\Omega^2 \\
		\sigma_{10} =54\sqrt{2}F_{1,1}'^* \Gamma^2 R_n V^* Z_T l\\
		\sigma_{11} =54\sqrt{2}F_{0,0}'^* \Gamma^2 R_n V^* Z_T l\\
		\sigma_{12} =54\sqrt{2}F_{1,1}'^* R_n V^* l\Omega^2 \\
		\sigma_{13} =54\sqrt{2}F_{0,0}'^* R_n V^* l\Omega^2 \\
		\sigma_{14} =54\sqrt{2}F_{1,1}'^* \Gamma^2 R_n V^* l\\
		\sigma_{15} =54\sqrt{2}F_{0,0}'^* \Gamma^2 R_n V^* l\\
		\sigma_{16} =54F_{1,1}^* R_n Z_T \Omega^2 \\
		\sigma_{17} =54F_{0,0}^* R_n Z_T \Omega^2 \\
		\sigma_{18} =54F_{1,1}^* \Gamma^2 R_n Z_T \\
		\sigma_{19} =54F_{0,0}^* \Gamma^2 R_n Z_T \\
		\sigma_{20} =54F_{1,1}^* R_n \Omega^2 \\
		\sigma_{21} =54F_{0,0}^* R_n \Omega^2 \\
		\sigma_{22} =54F_{1,1}^* \Gamma^2 R_n \\
		\sigma_{23} =54F_{0,0}^* \Gamma^2 R_n \\
		\sigma_{24} =4\Gamma +F_{0,0}^* R_n +F_{1,1}^* R_n +F_{0,0}^* R_n Z_T -F_{1,1}^* R_n Z_T \\+4F_{0,1}^* R_n V^*Z_T -\sigma_{33} -\sigma_{32} -\sigma_{31} +\sigma_{30} +2\\
		\sigma_{25} =\sqrt{2}F_{1,1}'^* \Gamma R_n V^*Z_T l\\
		\sigma_{26} =\sqrt{2}F_{1,1}'^* \Gamma R_n V^*Z_T l\\
		\sigma_{27} =\sqrt{2}F_{1,1}'^* \Gamma R_n V^* l\\
		\sigma_{28} =\sqrt{2}F_{1,1}'^* \Gamma R_n V^* l\\
		\sigma_{29} =2F_{0,1}^* \Gamma R_n V^*Z_T \\
		\sigma_{30} =\sqrt{2}F_{1,1}'^* R_n V^*Z_T l\\
		\sigma_{31} =\sqrt{2}F_{1,1}'^* R_n V^*Z_T l\\
		\sigma_{32} =\sqrt{2}F_{1,1}'^* R_n V^* l\\
		\sigma_{33} =\sqrt{2}F_{1,1}'^* R_n V^* l
	\end{gather*}
	\bibliography{ref-2}
\end{document}